\documentstyle[epsf]{elsart}

\begin{document}

\begin{frontmatter}

\title{\bf A Langevin equation for the energy cascade in 
fully-developed turbulence.}
\author[p]{Philippe Marcq}
\author[a]{and Antoine Naert}
\address[p]{Department of Physics, Graduate School of Science,\\
Kyoto University, Kyoto 606, Japan}
\address[a]{Laboratoire de Physique, Ecole Normale Sup\'erieure de
Lyon,\\
46 All\'ee d'Italie, 69394 Lyon C\'edex 07, France}

\begin{abstract}
Experimental data from a turbulent jet flow
is analysed in terms of an additive, continuous stochastic process
where the usual time variable is replaced by the scale.
We show that the energy transfer through scales  
is well described by a linear Langevin equation, and 
discuss the statistical properties of the corresponding
random force in detail. We find that
the autocorrelation function of the random force 
decays rapidly: the process is therefore Markov for scales larger
than Kolmogorov's dissipation scale $\eta$.
The corresponding autocorrelation scale is 
identified as the elementary step of the energy cascade.
However, the probability distribution function of the random force 
is both non-Gaussian and weakly scale-dependent.

\end{abstract}

\end{frontmatter}

\section{Introduction}
\label{sec-int}

Due to the molecular viscosity of the fluid, the kinetic energy of a flow
must be dissipated. The degree of instability of a flow is usually
quantified by the Reynolds number $Re$, defined as $Re = UL/\nu$
where $U$, $L$ and $\nu$ are respectively the mean velocity, the
largest scale of the flow, and the kinematic viscosity of the fluid. 
For large enough Reynolds number, the dissipation scale $\eta$ at which
energy is dissipated becomes much smaller than the integral scale $L$
at which it is injected: $L \gg \eta$.
The transport process through which energy injected at large scale is 
transported down to small scales is traditionally referred to as 
a ``cascade'', following Richardson's qualitative picture of turbulent eddies 
breaking down into smaller sub-eddies from the largest
scales of motion down to the dissipative scales \cite{book,review}.  

Richardson's ideas were first made quantitative by Kolmogorov \cite{K41}. 
Kol\-mo\-go\-rov's theory underlines the central role plaid by the 
mean energy dissipation rate $\langle \epsilon \rangle$.
Its main assumptions are that, for sufficiently large Reynolds number,
($i$) small scale turbulence (i.e. $r \ll L$) 
is homogeneous and isotropic, and thus decoupled from 
the large-scale properties of the flow, which may be 
affected by, e.g., the specific geometry of boundary
conditions; ($ii$) for scales large compared to the
dissipation scale ($r \gg \eta$), $\langle \epsilon \rangle$ is 
independent of $\nu$. Whenever ($i$) and ($ii$) hold,
one expects small scale statistical properties of the cascade process
to be universal, and determined only by $\langle \epsilon \rangle$
in the inertial range $\eta \ll r \ll L$.
In addition, dimensional arguments lead to a simple scaling form
for the probability distribution function of velocity increments.
This theory becomes inconsistent if one allows the 
energy dissipation rate to fluctuate.
However, experimental data shows unambiguously that the Kolmogorov 
picture is incomplete \cite{book,review}: the energy dissipation
rate $\epsilon$ is a strongly fluctuating quantity. Its 
inhomogeneous, intermittent behaviour 
effectively links small and large-scale features of the flow.
Further, the probability distribution function of 
ve\-lo\-ci\-ty in\-cre\-ments is not scale-invariant:
its shape is Gaussian at large scale, yet develops long
tails at smaller scale. 

The first phenomenological model to take into account the
strong variability of the energy dissipation rate  
was proposed by Kolmogorov and Obukhov \cite{KO62}. 
In this picture, emphasis is shifted from $\langle \epsilon \rangle$ 
to the local average
$\epsilon_r$ integrated over a ball of radius $r$. The fluctuations 
of the random variable $\ln \epsilon_r$ are assumed to be Gaussian:
\begin{equation}
\label{lognormal1}
P(\ln \epsilon_r,r) = {1 \over \Lambda(r) \sqrt{2 \pi}} \;
\exp \left( -{\left(\ln \epsilon_r - \langle \ln \epsilon_r
\rangle\right)^2 
\over 2 \Lambda^{2}(r) } \right).
\end{equation}
Further, the variance $\Lambda^2(r)$ 
depends logarithmically on scale:
\begin{equation}
\label{lognormal2}
\Lambda^2(r) = \Lambda_0^2 + \mu \ln (L/r),
\end{equation}
where $\mu$ is a constant and $\Lambda_0$ accounts for the large scale 
properties of the flow.
Log-normal probability distribution functions (Eq.~(\ref{lognormal1}))
are known to be a good first-order approximation of the statistics of
the 
energy transfer rate in fully-developed turbulent flows
\cite{logexp,logDNS}.
They arise naturally as solutions of 
multiplicative stochastic processes for the variable $\epsilon_r$
\cite{Yaglom,Mandelbrot},
where interaction is generally assumed to be local in scale:
the stochastic process must be Markov. However, this
assumption is generally not verified by experimental data
\cite{discrete}. 
Approaches related to the log-normal 
model (Eqs.~(\ref{lognormal1}-\ref{lognormal2}))
have attracted renewed attention recently 
\cite{Castaing90,Arneodo97}, despite well-known difficulties
linked to the scaling properties 
of $\epsilon_r$ and of the velocity increments $\delta v_r$.
\cite{Mandelbrot,nonlog}.

The statistical properties of isotropic and 
homogeneous turbulent flows have been reconsidered 
recently within the context of continuous, additive, Markov 
stochastic processes \cite{Friedrich97,Naert97}. Analysis of 
experimental data from a turbulent flow has shown that the
scale-dependence of probability distribution functions of 
velocity increments \cite{Friedrich97} 
and of the logarithm of the 
energy dissipation rate \cite{Naert97} may be described by 
Fokker-Planck equations, where the usual time variable is replaced by a 
monotonous function of scale $r$. 
The energy cascade  corresponds to an Ornstein-Uhlenbeck process 
for the stochastic variable $\ln \epsilon_r$,
since the drift and diffusion terms 
of the Fokker-Planck equation
are respectively found linear and constant.
This model is exactly solvable. 
The probability distribution function is found Gaussian: 
the statistics of $\epsilon_r$ are log-normal. 
However, the scale-dependence of the mean and variance 
differs from that considered in previously studied log-normal 
models \cite{KO62,Castaing90}.
The presence of a non-zero drift term implies that this new 
phenomenological approach is not equivalent to 
usual multiplicative models for the variable $\epsilon_r$.

In this work, we consider an alternative
formulation of the model introduced in \cite{Naert97}.
Our analysis is based on the same velocity signal, recorded 
in a low-temperature turbulent jet flow at high Reynolds number 
$Re = 20000$. We aim at building a faithful, yet simple model of 
available experimental data, from a phenomenological perspective. 
Fokker-Planck equations, which govern the evolution of
probability distribution functions, are equivalent 
mathematically to stochastic differential equations
with an additive noise term (Langevin equations, see \cite{Risken}).
The goal pursued here is to investigate the energy
transfer process from the complementary viewpoint
given by stochastic trajectories in scale. 
A detailed study of statistical quantifiers which arise naturally
in this new context allows to check the overall consistency of the
approach. In particular, the validity of approximations 
made in \cite{Naert97} is investigated in detail. 

This article is organised as follows. The experimental
set-up is discussed in Sec.~\ref{sec-exp}, together 
with the specifics of data acquisition and signal processing.
Prescriptions used in the analysis of the experimental velocity
signal are introduced in Sec.~\ref{sec-data}. 
The results previously obtained in \cite{Naert97} 
are summarised in Sec.~\ref{sec-stoc}. The stochastic cascade process 
is next examined in detail within the context of a Langevin description
in Sec.~\ref{sec-Langevin}. The Markov and Gaussian nature of the
process are then discussed in Secs.~\ref{sec-Markov} and \ref{sec-stat}
respectively. Sec.~\ref{sec-conc} is devoted
to a summary and discussion of our results.

\section{A low-temperature jet flow}
\label{sec-exp}

The velocity signal analysed in this work was recorded in a 
low-temperature gaseous helium jet flow 
\cite{Chabaud94,Castaing94,Naert95}. Experimental conditions 
(temperature $T=4.2$ K, pressure $P = 1$ bar) 
are set close to the liquid-gas transition: 
this ensures a very low kinematic viscosity, 
$\nu \simeq 2 \times 10^{-8}$ 
m$^2$ s$^{-1}$, and thus gives access to high Reynolds numbers, up 
to $Re \sim 10^5$ in this cell. 

After laminarisation, gaseous helium is injected through a contracting
nozzle (diameter $2$ mm) into a cylindrical cell (diameter $13$ cm,
height $30$ cm) where the turbulent jet develops. 
A grid is set $17$ cm downstream from the nozzle to avoid  
instability occurring on long time scales due 
to recirculation of the flow along the walls.
Further downstream, gaseous helium is pumped out of the cryostat.
Adjusting the microvalve inlet opening and the outlet pumping 
allows to vary not only the flow rate (mean velocity $U$), but also 
the operating pressure $P$ in the cell. Since the product $P \nu$
is roughly constant, this original feature of the experiment allows
to vary the Reynolds number over the range $10^3 \le Re \le 10^5$.
Access to a large range of Reynolds number for a fixed geometry
is the  main benefit of this cryogenic experiment
(see \cite{Naert95} for further details). 

The longitudinal component of the velocity is measured
thanks to a cryogenic hot-wire anemometer \cite{Castaing94}
located $10$ cm downstream from the nozzle on the axis of the jet, 
where the flow is fully turbulent.
The effective spatial resolution is estimated to be $22 \mu$m. 
The wire is operated at constant resistance through a home-made
$10$ MHz lock-in amplifier.
The signal is next digitalised using a $12$ bits A/D converter. 
We believe that the dominant source of experimental error is 
the resolution of the digitalization. 

What represents, from a physical point of view, the signal recorded 
by the hot wire? The velocity field is a function of space and time
$v(x,t)$.
In a jet geometry, as in many experimental flows, the turbulent 
velocity field is advected relatively to the observer (anemometer).
The ``rate of velocity fluctuations'' 
is ususally defined as the ratio of the standard deviation of the
velocity 
over its mean value. 
A small value of this rate means that the flow is effectively
``frozen'': Taylor's frozen turbulence hypothesis is valid \cite{book}.
Rapid advection of the velocity field $v(x,t)$ with 
respect to the (fixed) detector ensures that the signal
records the spatial structure of the velocity field ($v(x,t) \simeq
v(x)$,
with $x = \langle v \rangle t$), rather than its temporal evolution.
In this jet, the rate of fluctuation is $28 \%$, in agreement 
with values typical of jet flows. 
The frozen turbulence hypothesis is now a source of error \cite{HDT1},
in fact 
a bias on estimates of the spatial velocity field $v(x)$. 

In the present work, this point is taken into account
by resampling the signal as follows 
(see \cite{HDT2} for further details).
The spatial coordinate $x$ of a record taken at 
time $t$ is $x(t)=\int_{0}^{t}v(t)dt$. The bias on the velocity 
signal can be eliminated by considering instead the series $v(x(t))$. 
However, the interval between two successive points becomes irregularly 
distributed. One must also resample the signal regularly, in practice 
by linear interpolation. We checked that a more precise interpolation, 
involving the second order derivative, does not sensibly improve
the regularity of sampling, and leaves our measurements unchanged.
Moreover, our conclusions do not seem to depend on 
the choice of a specific correction method, since
results remain unchanged when using, e.g., the local Taylor hypothesis 
advocated in \cite{HDT1}. 

For consistency, the velocity signal we analyse is the same as studied 
in \cite{Naert97}, and was recorded at $Re=20 000$. The Reynolds number 
based on Taylor's microscale
is $R_\lambda=341$. The dissipation scale is estimated 
to $\eta = 20 \mu$m according to Kolmogorov's formula:
\begin{equation}
\label{defeta}
\eta =  {\left({\nu^3}\over{\langle \epsilon \rangle}\right)}^{1/4},
\end{equation}
with $\langle \epsilon \rangle$ defined as 
$15 \nu \; \langle \; \left( {\rm d}v / {\rm d}x \right)^2 \; \rangle$.
The smallest structure of the flow is therefore resolved at this value 
of $Re$: this justifies our choice of this specific sample.
The integral scale, operationally defined as the correlation 
length of the velocity signal, 
is evaluated to $L \simeq 1$ cm ($L \simeq 500 \eta$).
This numerical value indeed corresponds to the largest structure of 
the flow, since the diameter of the jet is about $2.5$ cm
in the vicinity of the anemometer, where the measurement is done.
Note however that the values of
$\eta$ and $L$ only have qualitative importance in our study.

\section{Prescriptions for data analysis}
\label{sec-data}

As proposed by Obukhov \cite{KO62}, the dissipation averaged 
over a volume of size $r$ may be used
as a reasonable Ansatz for the local transfer rate of energy at scale
$r$. 
Since available experimental data is often limited to the 
longitudinal velocity component $v(x)$,
it is customary to use the following one-dimensional
surrogate of the dissipation $\epsilon_r(x)$ at location $x$:
\begin{equation}
\label{epsilon}
\epsilon_r(x) =
\frac{15\nu}{r} \int_{x-r/2}^{x+r/2}
\left( \frac{dv}{dx'} \right)^2 dx',
\end{equation}
a definition which we endorse here.
Eq.~(\ref{epsilon}) is generally believed to capture
most of the relevant physics (see \cite{review} for
up-to-date critical overviews).

Even though the relevant physics is continuous, experimental data 
has been sampled before being recorded: the available data is thus
discrete, with a sampling distance equal to 
$d = 22 \mu$m (or $d = 1.1 \eta$).
In what follows, the unit of length
is set to $d$ unless otherwise noted. 
Values of the dissipation field $\epsilon_r(x)$
are obtained thanks to a first-order finite-difference approximation of
the derivative of the velocity field:
\begin{equation}
\label{ediscrete}
\epsilon_r(x) =
\frac{15\nu}{r} \sum_{x'=x-{\rm int}(r/2)}^{x+{\rm int}(r/2)-1}
\left( {v \left( x'+\delta x' \right) - v(x') \over \delta x'}
\right)^2,
\end{equation}
where $\delta x'$ is the (constant) discretisation step.
The numerical value:
\begin{equation}
\label{choice1}
\delta x' = 5,
\end{equation}
is adopted 
in all calculations reported here.
Note that with this choice, the fluctuations of $\epsilon_r$ at 
$r \simeq \eta$ cannot be resolved since $\delta x' = 5.5 \eta$.

In the following, we will refer to both variables $r$ and $l$
as ``scale'', where the new variable $l$ is related to the physical
scale $r$ 
(as used in Eq.~(\ref{epsilon})) by :
\begin{equation}
\label{defl}
l = \ln\left(L \over r \right).
\end{equation}
This definition is used for convenience since it
leads to simpler analytic expressions.
For the relevant scales $\eta \le r \le L$, the continuous variable $l$ 
is positive and bounded from above:
\begin{equation}
0 \le l \le \ln\left(L \over \eta \right)
\end{equation}
It is a monotonously decreasing function of the
usual scale $r$. 
For clarity, scale will however be measured in 
units of the dissipation scale $\eta$ in most graphs.

The random process $X(l)$ is defined as:
\begin{equation}
X(l) = \ln \epsilon_r,
\end{equation}
using Eqs.~(\ref{epsilon}) and (\ref{defl}).
Statistical averages, denoted by the symbol $\langle \; \rangle$,
are evaluated over the whole data set ($10^7$ points, or about $2\;
10^4L$).
For all statistical quantifiers presented here,
this set is large enough to ensure statistical convergence,
as confirmed by comparisons with results obtained from smaller samples.

\begin{figure}[thb]
\vspace{-0.5cm}
\centerline{
\epsfxsize 10cm
\epsffile{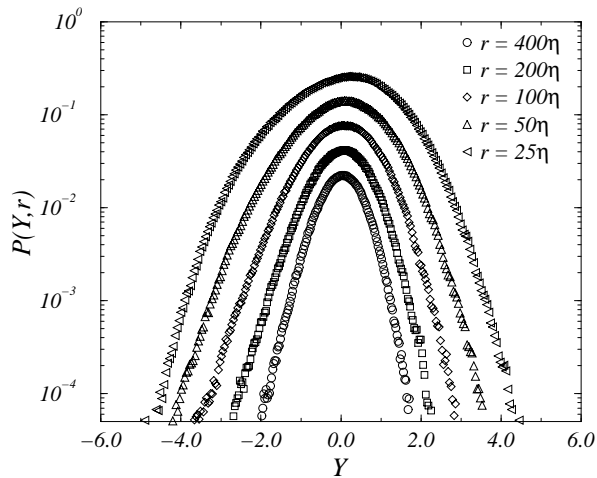}
}
\vspace{-0.5cm}
\caption{Probability distribution functions $P(Y,r)$ 
of the centered variable Y at scales $r$ ranging from 
$r = 25 \eta$ to $400 \eta$. The variance $\langle Y(r)^2 \rangle$
is a decreasing function of scale. Asymmetry of the histograms 
translates into a non-zero value of the skewness. For clarity,
histograms were shifted by a factor $2$ along the vertical axis.}
\label{fig-pdfy}
\end{figure}

For convenience, we will also consider
the centered variable $Y(l)$, defined by:
\begin{equation}
\label{defY}
Y(l)=X(l)-\langle X(l) \rangle.
\end{equation}
The probability distribution functions $P(Y,r)$ of the variable $Y(r)$
computed according to Eq.~(\ref{ediscrete}) are
shown in Fig.~\ref{fig-pdfy}. Despite rough agreement
with the parabolic shape expected for a Gaussian distribution
in a lin-log plot, an obvious departure from the functional form 
(\ref{lognormal1}) is the asymmetry of $P(Y,r)$. This point has already
been noted in previous experimental \cite{logexp} and numerical work 
\cite{logDNS}.

Evaluation of the scale-derivative of $Y(l)$ will also be needed
in the following. A first-order approximation will be used:
\begin{equation}
\label{dYdl}
{{\rm d}Y \over {\rm d}l}(l) = {1 \over \delta l} \;
\left( Y\left(l+\delta l \right) - Y(l) \right);
\end{equation}
where the step $\delta l$ is naturally related to the physical scale
difference
$\delta r$ by the relation:
\begin{equation}
\label{dldiscrete}
\delta l(r) = \ln \left( {r + \delta r \over r} \right).
\end{equation}
We use here the numerical value:
\begin{equation}
\label{choice2}
\delta r = 4,
\end{equation}
or $\delta r = 4.4 \eta$.
The discretisation step is thus a decreasing function of physical scale
$r$: $\delta l(r = 25 \eta) \sim 0.17$,
$\delta l(r = 100 \eta) \sim 0.04$, $\delta l(L) \sim 0.009$.
Our results are unchanged 
for other reasonable numerical values of the discretisation steps 
(\ref{choice1}) and (\ref{choice2}). This ensures that the continuous
limit is well-controlled.  

\section{An Ornstein-Uhlenbeck process}
\label{sec-stoc}

In this section, we wish to summarise the analysis made in
\cite{Naert97} of the stochastic process $X(l)$, as defined above.

A first important result is that the
transition probability distribution function 
$P\left(X',l'|X,l\right)$ $ = P\left(X(l')|X(l)\right)$
respects the Chapman-Kolmogorov equation to a very good approximation.
Mathematically, the Chapman-Kolmogorov equation is a 
necessary (but not sufficient) condition for the process to be Markov.
An equivalent (differential) formulation is the Kramers-Moyal 
expansion \cite{Risken}, an evolution equation for
the probability distribution function $P(X,l)$ which reads:
\begin{equation}
\label{KM}
\frac{\partial}{\partial l} P(X,l) = 
\sum_{n=1}^{\infty} (-1)^n  \frac{\partial^n}{\partial
X^n} \left( D_n(X,l)  P(X,l) \right).
\end{equation}
The Kramers-Moyal coefficients $D_n(X,l)$ in Eq.~(\ref{KM}) are defined
as the limit:
\begin{equation}
\label{KM1}
D_n(X,l) = \lim_{l' \rightarrow l} M_n\left(X,l,l'\right),
\end{equation}
where the functions $M_n\left(X,l,l'\right)$ are
equal, for $l \neq l'$ and up to a multiplicative factor,
to the $n$-th order moment of the transition probability
distribution function:
\begin{equation}
\label{KM2}
M_n\left(X,l,l'\right) = \frac{1}{l-l'} \; \frac{1}{n!} \;
\int (X'-X)^n \; P(X',l'|X,l) \; dX'.
\end{equation}

In \cite{Naert97}, the coefficients $D_n(X,l)$ are equated 
to numerical values found for $M_n(X,l,l'=l+ \delta l)$, 
$\delta l = 0.04$, below which resolution becomes insufficient. 
According to this procedure, the second and fourth-order 
coefficients $D_2$ and $D_4$ are found roughly constant, 
i.e. independent of both variables $X$ and $l$, with
$D_4 \le 0.05 (D_2)^2$, and $D_2= 0.03 \pm 0.005$. 
The inequality is interpreted
by neglecting the fourth-order term in Eq.~(\ref{KM}): $D_4(X,l) = 0$.
If $D_4 = 0$, Pawula's theorem  \cite{Risken} then implies that the
expansion 
(\ref{KM}) may be truncated at second order.
In particular, $D_3(X,l) = 0$, 
and Eq.~(\ref{KM}) reduces to the 
Fokker-Planck equation:
\begin{equation}
\label{FP}
\frac{\partial}{\partial l} P(X,l) = - \frac {\partial}{\partial X} 
\left(D_1(X,l) P(X,l) \right)+ \frac{\partial^2}{\partial X^2}
\left(D_2(X,l) P(X,l) \right),
\end{equation}
where $D_2(X,l) = D$ is a diffusion constant.

Using the same procedure, the first-order coefficient is found linear
in $X$, with an approximately scale-independent slope $\gamma$.
Using at each scale the centered variable $X-\langle X(l) \rangle$,
one may write:
\begin{equation}
\label{drift}
D_1(X,l) = \gamma (X-\langle X(l) \rangle) + F(l),
\end{equation}
where $F(l)$ is a (real) function of scale, and $\gamma = 0.21 \pm
0.02$.
Eqs.~(\ref{FP})-(\ref{drift}) define 
an Ornstein-Uhlenbeck process with drift and diffusion 
coefficients $\gamma$ and $D$ \cite{Risken}.
The drift coefficient $\gamma$, a real  number, should
not be confused with the drift {\it term} $D_1(X,l)$, a real
function of variables $X$ and $l$.
Note that $\gamma$ is found positive: the process does not relax 
to a stationary state for large $l$. 

In \cite{Naert97}, Eq.~(\ref{FP}) is integrated by assuming
that energy does not fluctuate at the largest scale $L$, as 
expected in the ideal case of an ensemble of systems 
supplied with the same power. The corresponding 
initial condition reads:
\begin{equation}
\label{CI}
P(X,l=0)= \delta \left(X(0)- \ln \epsilon_L \right)
\end{equation}
The Fokker-Planck equation supplemented with Eq.~(\ref{CI}) 
admits an exact, Gaussian solution:
\begin{equation}
\label{Gauss}
P(X,l) = {1 \over \Lambda(l) \sqrt{2 \pi}} \;
\exp \left( -{(X - \langle X(l) \rangle)^2 \over 2 \Lambda^{2}(l) }
\right).
\end{equation}
Similarly to Kolmogorov and Obukhov's model \cite{KO62}, the dissipation 
exhibits log-normal fluctuations. 

Since energy is conserved through the cascade 
(i.e. $\forall r \in [\eta, L]$,
$\langle \epsilon_r \rangle = \langle \epsilon \rangle$),
the mean and variance of the Gaussian probability distribution
function (\ref{Gauss}) are simply related by:
\begin{equation}
\label{xmean}
\langle X(l) \rangle = \langle X(0) \rangle - \Lambda^{2}(l) / 2,
\end{equation}
in good agreement with experimental data.
According to Eq.~(\ref{CI}), we postulate that
$\langle X(0) \rangle = X(0) = \ln \epsilon_L$.
The variance $\Lambda^2(l)$ of the Gaussian solution 
(\ref{Gauss}) is a monotonously decreasing function of scale $r$:
\begin{equation}
\label{lambda}
\Lambda^{2}(l) = {D \over \gamma} \left( e^{2\gamma l}-1\right) = 
{D \over \gamma} \left( \left({L \over r}\right)^{2\gamma}-1\right).
\end{equation}
A consistency condition is:
\begin{equation}
\label{defF}
F(l) = {d  \over dl} \langle X(l) \rangle = - D e^{2\gamma l},
\end{equation}
in good agreement with direct measurements of the first-order
Kramers-Moyal coefficient $D_1(X,l)$.
Note that Eq.~(\ref{lognormal2}) is recovered when the drift coefficient
$\gamma$ cancels, with $\mu = 2 D$ and $\Lambda_0^2 = 0$.
For small enough scales ($r \ll L$), Eq.~(\ref{lambda}) reduces
to a power-law dependence, in agreement with Castaing's model.
In that context, $\gamma$ was observed to be inversely proportionnal 
to $\log(Re)$ \cite{Castaing90,Chabaud94}. 

One may summarise the main result of \cite{Naert97} as follows:
an Ornstein-Uhlenbeck process with positive drift coefficient
(Eqs.~(\ref{FP})-(\ref{drift})) gives a 
simple picture of the scale dependence of energy transfer statistics
of a turbulent flow. The validity of this description
relies on two assumptions: first that the process is Markov, 
second that the Kramers-Moyal
coefficients of order strictly larger than $2$ may be neglected,
in particular that $D_3(X,l)$ and $D_4(X,l)$ may be set to zero.
The analysis made in \cite{Naert97} suggests that both assumptions are
reasonably well-supported by experimental data. 
However, one may already note that the non-zero, negative skewness 
of $P(X,l)$ observed in Fig.~\ref{fig-pdfy} is not taken into 
account by the Fokker-Planck model.
In what follows, we adopt the complementary perspective provided 
by a formally equivalent Langevin description in order to better 
assess the validity of these approximations.

\section{A Langevin equation}
\label{sec-Langevin}

We will study in this section the energy cascade process
from the perspective given by scale-dependent random
trajectories of the centered stochastic variable $Y(l)$.
A Langevin equation formally equivalent to the Fokker-Planck
equation (\ref{FP}) is first introduced in 
Sec.~\ref{sec-Langevin-previous}.
We proceed to measure directly the drift
and diffusion coefficients $\gamma$ and $D$ 
(Secs.~\ref{sec-Langevin-mes}) by methods which do not involve 
the computation of Kramers-Moyal coefficients.
Predictions involving the two-point correlators 
of $Y(l)$ are next confronted to experimental data 
in Sec.~\ref{sec-Langevin-pred}. A definition of
the random force is finally introduced in 
Sec.~\ref{sec-Langevin-disc}, thanks to proper
discretisation of the Langevin equation.

\subsection{The equivalent Langevin description}
\label{sec-Langevin-previous}

Eqs.~(\ref{FP}-\ref{CI}) and (\ref{defF}) are mathematically equivalent
to
the stochastic differential equation \cite{Risken}:
\begin{equation}
\label{Langevin}
{dY \over dl}(l) = \gamma Y(l) + \sqrt{2D} \; \xi_{\rm FP}(l)
\end{equation}
with initial condition $Y(0) = 0$ (note that 
$F(l) = d\langle X(l) \rangle /dl$).
When strict equivalence with the Fokker-Planck equation (\ref{FP})
holds, the random force $\xi_{\rm FP}(l)$ 
is a stationary, Gaussian, white noise and the stochastic
process is Markov. The two-point correlation function therefore reads:
\begin{equation}
\label{noise2}
\left\langle \xi_{\rm FP}(l) \; \xi_{\rm FP}(l') \right\rangle =
\delta(l-l'),
\end{equation}
and the probability distribution function is a scale-independent
Gaussian:
\begin{equation}
\label{noise3}
P(\xi_{\rm FP}, l) =  { 1 \over \sqrt{2 \pi}} \;
\exp \left( -{\left. \xi_{\rm FP}\right. ^2 \over 2} \right),
\end{equation}
with mean $\langle \xi_{\rm FP}(l) \rangle = 0$
and variance $\langle \xi_{\rm FP}(l)^2 \rangle = 1$.

The linear equation (\ref{Langevin}) is exactly solvable.
Its solution reads:
\begin{equation}
\label{solution1}
Y(l)  = Y(0) + \sqrt{2 D} \int_0^l{e^{\gamma (l - l')} \; \xi(l') \; 
{\rm d}l'}.
\end{equation}
Two-point statistics of the process $Y(l)$ can also be calculated
when the random force is delta-correlated (Eq.~(\ref{noise2})).
Assuming that the initial distribution $P(Y,0)$ admits a finite
width $\langle Y(0)^2 \rangle$, one obtains:
\begin{equation}
\label{varY}
\langle Y(l)^2 \rangle  = \langle Y(0)^2 \rangle  + 
{D \over \gamma} \left( e^{2 \gamma l} - 1 \right).
\end{equation}
Expression (\ref{lambda}) is recovered when $\langle Y(0)^2 \rangle =
0$,
as expected. Further, we define the normalised scale autocorrelation 
function of the stochastic process $Y(l)$ by the relation:
\begin{equation}
\label{def-cory}
C_Y(l,\Delta l) = {\langle Y(l) \; Y(l+\Delta l) \rangle
\over \langle Y(l)^2 \rangle},
\end{equation}
where $\Delta l$ is a positive scale difference.
A straightforward calculation leads to:
\begin{equation}
\label{corY}
\ln C_Y(l,\Delta l) = \gamma \; \Delta l.
\end{equation}
Note that expressions (\ref{varY}) and (\ref{corY}) are obtained 
independently of the functional form of the probability distribution
function of the random force.

\subsection{Drift and diffusion coefficients}
\label{sec-Langevin-mes}

\begin{figure}[thb]
\vspace{-0.5cm}
\centerline{
\epsfxsize 10cm
\epsffile{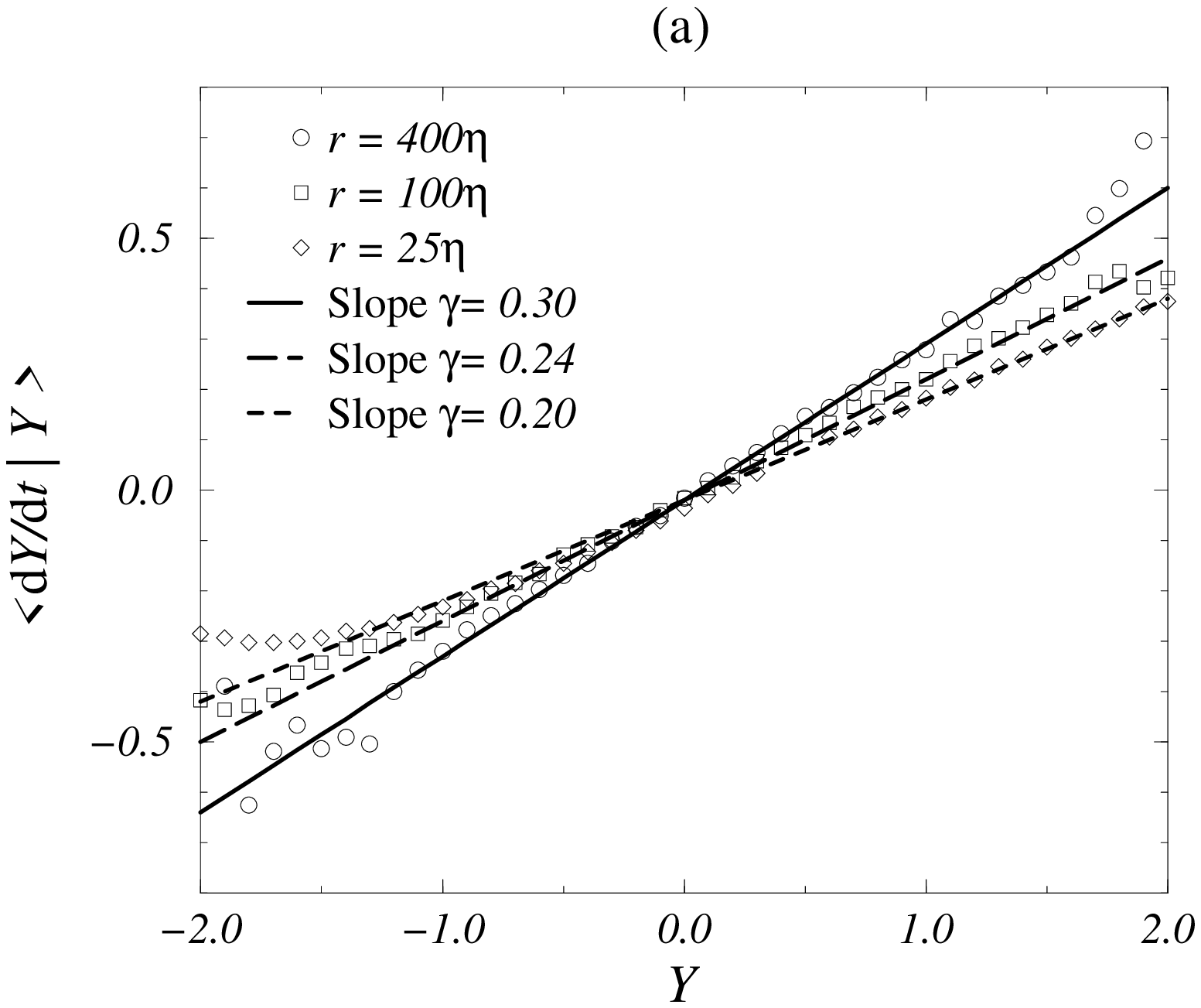}
\hspace{-1.5cm}
\epsfxsize 10cm
\epsffile{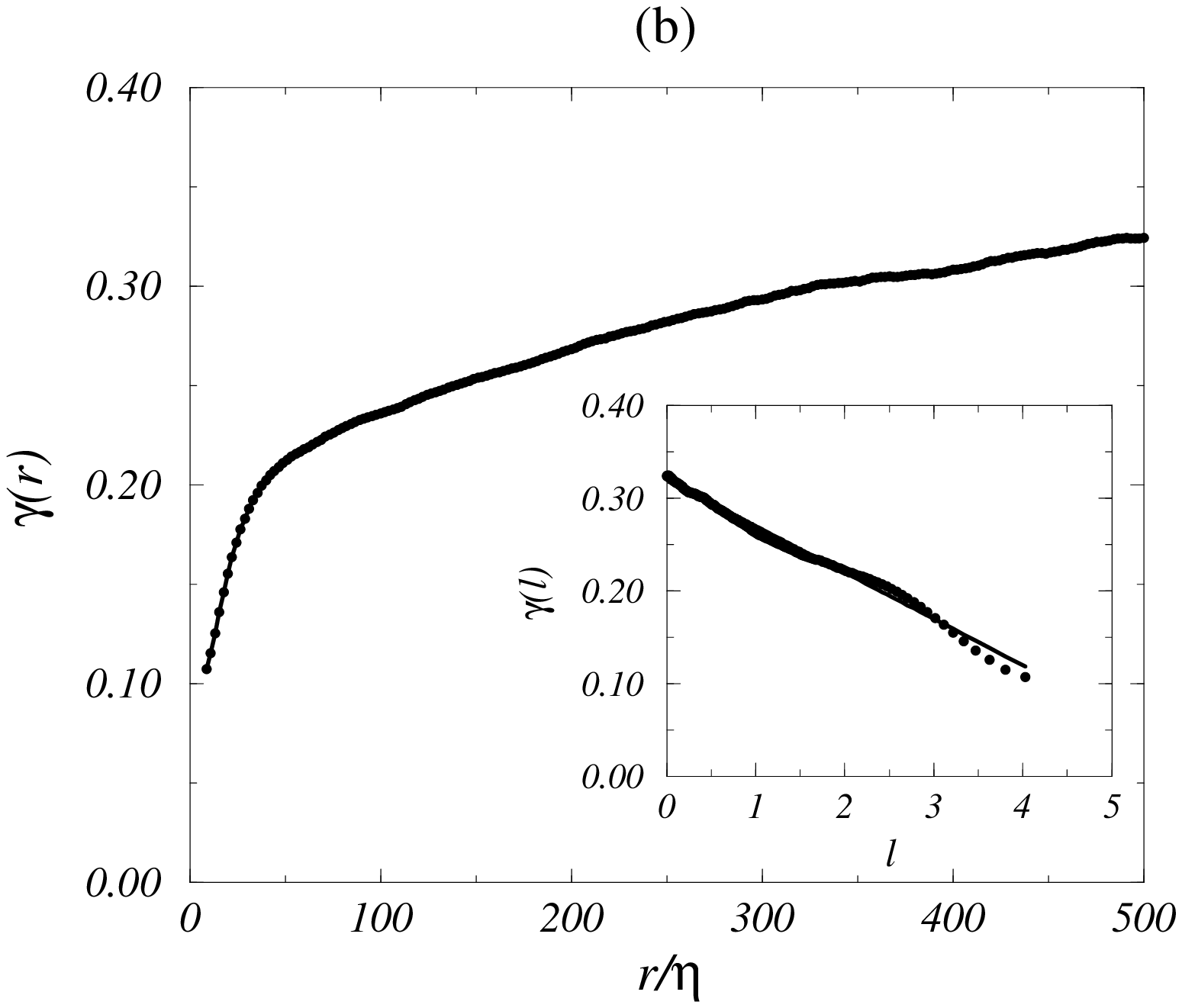}
}
\vspace{-0.5cm}
\caption{Measure of the drift coefficient $\gamma(l)$. 
Graph (a): plot of the conditional average 
$\langle \dot{Y}(l) | Y(l) \rangle$ 
vs. $Y(l)$, estimated at scales $r = 25 \eta, 100 \eta$ and $400 \eta$.
Straight lines of slope $0.20$, $0.24$ and $0.30$ are drawn
to guide the eye.
Graph (b): plot of $\gamma(r)$, as defined in Eq.~(\ref{mesgam2}),
versus scale $r/\eta$. Statistical error is of
the order of a few percent in relative value. 
Inset: log-lin plot of the same data ($l = \ln(L/r)$).
The straight line corresponds to $\gamma(l) = 0.32 - 0.05 \; l$.}
\label{fig-gamma}
\end{figure}

Assuming that the energy cascade process can be described
by a Langevin equation such as (\ref{Langevin}), one would
first of all like to check whether this equation is indeed linear.
Since $\langle \xi(l) \rangle = 0$, a straightforward 
consequence of Eq.~(\ref{Langevin}) is:
\begin{equation}
\label{mesgam1}
\left\langle \dot{Y}(l) | Y(l) \right\rangle = \gamma Y(l),
\end{equation}
where $\dot{Y}(l)$ denotes the derivative of $Y(l)$ with respect
to scale. It is easy to see that nonlinear functions of $Y(l)$ 
in the right hand side 
of Eq.~(\ref{Langevin}) would also contribute
to the the conditional average $\langle \dot{Y}(l) | Y(l) \rangle$.
Direct measures of $\langle \dot{Y}(l) | Y(l) \rangle$ are presented
in Fig.~\ref{fig-gamma}, where the prescriptions detailed 
in Sec.~\ref{sec-data} were used. Within statistical
error, the conditional average $\langle \dot{Y}(l) | Y(l) \rangle$
is indeed proportional to $Y(l)$, even though
a small, possibly cubic contribution appears at small scales
for (rare) large negative values of the variable $Y(l)$.
This observation confirms the linearity of the relevant Langevin 
equation. However, the corresponding slope $\gamma$ seems to be
an increasing function of scale.

Using Eq.~(\ref{mesgam1}) to obtain quantitatively accurate
estimates of $\gamma(l)$ is rather costly numerically. 
We will turn to a simpler method, which turns out to yield
consistent estimates. Implicit in Eq.~(\ref{Langevin})
is the assumption that the random force $\xi_{\rm FP}(l)$ is 
statistically independent from the stochastic variable $Y(l)$ 
at all scales $l$:
\begin{equation}
\label{noise1}
\left\langle Y(l) \; \xi_{\rm FP}(l) \right\rangle = 0.
\end{equation}
Upon multiplying both sides of the Langevin equation
(\ref{Langevin}) by $Y(l)$, ensemble-averaging, and using
Eq.~(\ref{noise1}), one easily obtains:
\begin{equation}
\label{mesgam2}
\gamma(l) = {\langle \dot{Y}(l) \; Y(l) \rangle \over
\langle Y(l)^2 \rangle},
\end{equation}
where the drift coefficient $\gamma(l)$ is a priori a function of scale.
Fig.~\ref{fig-gamma}.b shows that $\gamma(l)$, as estimated thanks 
to Eq.~(\ref{mesgam2}), is indeed a slowly varying function of scale.
The order of magnitude is however the same as that of
the (constant) value advocated in \cite{Naert97} ($0.21 \pm 0.02$).
This method appears to be more sensitive than direct calculations
of the Kramers-Moyal coefficients.

We would like to emphasise that: ($i$) Eqs.~(\ref{mesgam1}) and 
(\ref{mesgam2}) yield mutually consistent estimates of $\gamma(l)$;
($ii$) 
the normalised cross-correlation function $C_{Y \xi}(r)$, defined as:
\begin{equation}
\label{coryxi}
C_{Y \xi}(r) = {\left\langle Y(r) \; \xi(r) \right\rangle \over
Y_{\rm rms}(r) \; \xi_{\rm rms}(r)},
\end{equation}
is indeed close to zero (in practice of the order of $10^{-2}$)
when the random force $\xi(l)$ is calculated thanks to Eq.~(\ref{bruit})
for constant drift and diffusion coefficients, e.g. 
$\gamma(l) = \gamma = 0.21$ and $D(l) = D = 0.03$. 

As shown in the inset of Fig.~\ref{fig-gamma}, the scale dependence 
of the drift coefficient is well described by
a linear function of $l$ over the whole range
of relevant scales. Introducing a factor $2$ for convenience,
we write:
\begin{equation}
\label{valgam}
\gamma(l) = \gamma_0 - 2 \gamma_1 l.
\end{equation}
Our estimates of the constants $\gamma_0$ and $\gamma_1$ are
$\gamma_0 = 0.32 \pm 0.03$ and $\gamma_1 = 0.025 \pm 0.003$.
The error bars take into account both statistical error and 
the uncertainty deriving from the existence of other possible
choices of the discretisation steps $\delta r$
and $\delta x'$. Note that the drift coefficient is positive
for all physically relevant scales: Eq.~(\ref{valgam}) yields
$\gamma(r = \eta) = 0.01 > 0$.

\begin{figure}[thb]
\vspace{-0.5cm}
\centerline{
\epsfxsize 10cm
\epsffile{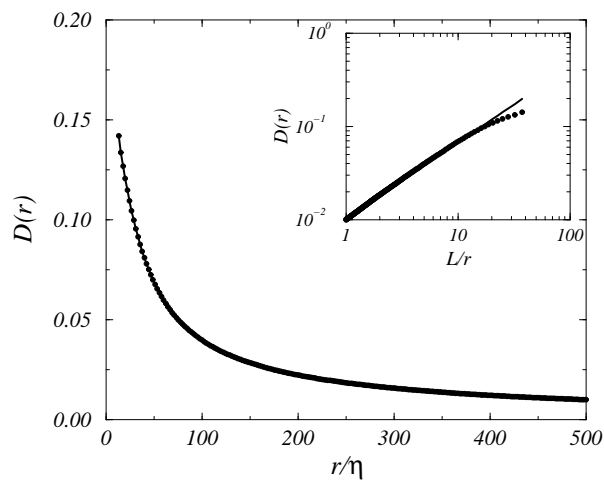}
}
\vspace{-0.5cm}
\caption{Graph of the diffusion coefficient $D(l)$ (as defined in
Eq.~(\ref{mesD})) versus scale $r/\eta$. Statistical errors are smaller
than the width of the symbols used. Inset: same data in log-log scale. 
The straight line corresponds to the scaling law
$D(r) = 0.01 (L/r)^{0.4}$.}
\label{fig-D}
\end{figure}

Assume now that the second-order
Kramers-Moyal coefficient $D_2(X,l)$ is independent of $X$,
and reduces to a (possibly scale-dependent) diffusion 
coefficient $D(l)$. According to Eq.~(\ref{Langevin}), this
coefficient may then be defined as:
\begin{equation}
\label{mesD}
2 D(l) = \left\langle \left({dY \over dl}(l) - \gamma(l) \; Y(l)
\right)^2 \right\rangle,
\end{equation}
where the variance of the random force is set to unity:
$\langle \xi(l)^2 \rangle = 1$.
Using Eq.~(\ref{mesD}), where $\gamma(l)$ is evaluated
thanks to Eq.~(\ref{mesgam2}), we find a finite, 
{\it scale-dependent}, monotonously 
decreasing coefficient diffusion $D(r)$ (as a function of
the physical scale $r$, see Fig.~\ref{fig-D}). 
The order of magnitude
is consistent with that of the constant value advocated in 
\cite{Naert97} ($D = 0.03 \pm 0.01$).
An excellent fit of the data is given by the expression:
\begin{equation}
\label{valD}
D(l) = D_0 \; e^{2 \; \delta \; l},
\end{equation}
with $D_0 = 0.01 \pm 0.005$ and $\delta = 0.40 \pm 0.01$
(see the inset of Fig.~\ref{fig-D}), except at scales $r$
close to the dissipation scale $\eta$.

We have shown in this section that the drift and diffusion
coefficients, when measured according 
to Eqs.~(\ref{mesgam2}) and (\ref{mesD}), are in fact not independent 
of the scale $l$. Since this scale-dependence cannot be 
attributed to, say, statistical errors, it becomes essential to 
understand to what extent the coefficients $\gamma(l)$ and $D(l)$ may be
approximated by constant coefficients, as has been advocated
in \cite{Naert97}. This is the goal of the next section,
where predictions concerning the two-point statistics
of $Y(l)$ will be confronted to experimental data
for both scale-dependent and constant coefficients.

\subsection{Two predictions}
\label{sec-Langevin-pred}

Instead of Eq.~(\ref{Langevin}), we
now use the following Langevin equation as our starting point:
\begin{equation}
\label{Langevin2}
{dY \over dl}(l) = \gamma(l) Y(l) + \sqrt{2 D(l)} \; \xi(l),
\end{equation}
where the functions $\gamma(l)$ and $D(l)$ are
given by Eqs.~(\ref{valgam}) and (\ref{valD}) respectively.
The solution of Eq.~(\ref{Langevin2}) reads:
\begin{equation}
\label{solution2}
Y(l) = e^{\gamma_0 l - \gamma_1 l^2} \left( Y(0) +
\sqrt{2 D_0} \int_0^l{e^{(\delta - \gamma_0) l' + \gamma_1 l'^2}
\xi(l') {\rm d}l'} \right) .
\end{equation}
Assuming that the random force $\xi(l)$ is delta-correlated, we obtain 
the following expression of the variance of $Y(l)$:
\begin{equation}
\label{varY2}
\langle Y(l)^2 \rangle = e^{2 \gamma_0 l - 2 \gamma_1 l^2} 
\left( \langle Y(0)^2 \rangle +
2 D_0 \int_0^l{e^{2 (\delta - \gamma_0) l' + 2 \gamma_1 l'^2} {\rm d}l'}
\right)
\end{equation}
for an initial width $\langle Y(0)^2 \rangle$.
The normalised scale autocorrelation function of the stochastic
process $Y(l)$ then reads:
\begin{equation}
\label{corY2}
\ln C_Y(l,\Delta l) = (\gamma_0 - 2 \gamma_1 l) \Delta l - \gamma_1 
(\Delta l)^2.
\end{equation}

\begin{figure}[thb]
\vspace{-0.5cm}
\centerline{
\epsfxsize 10cm
\epsffile{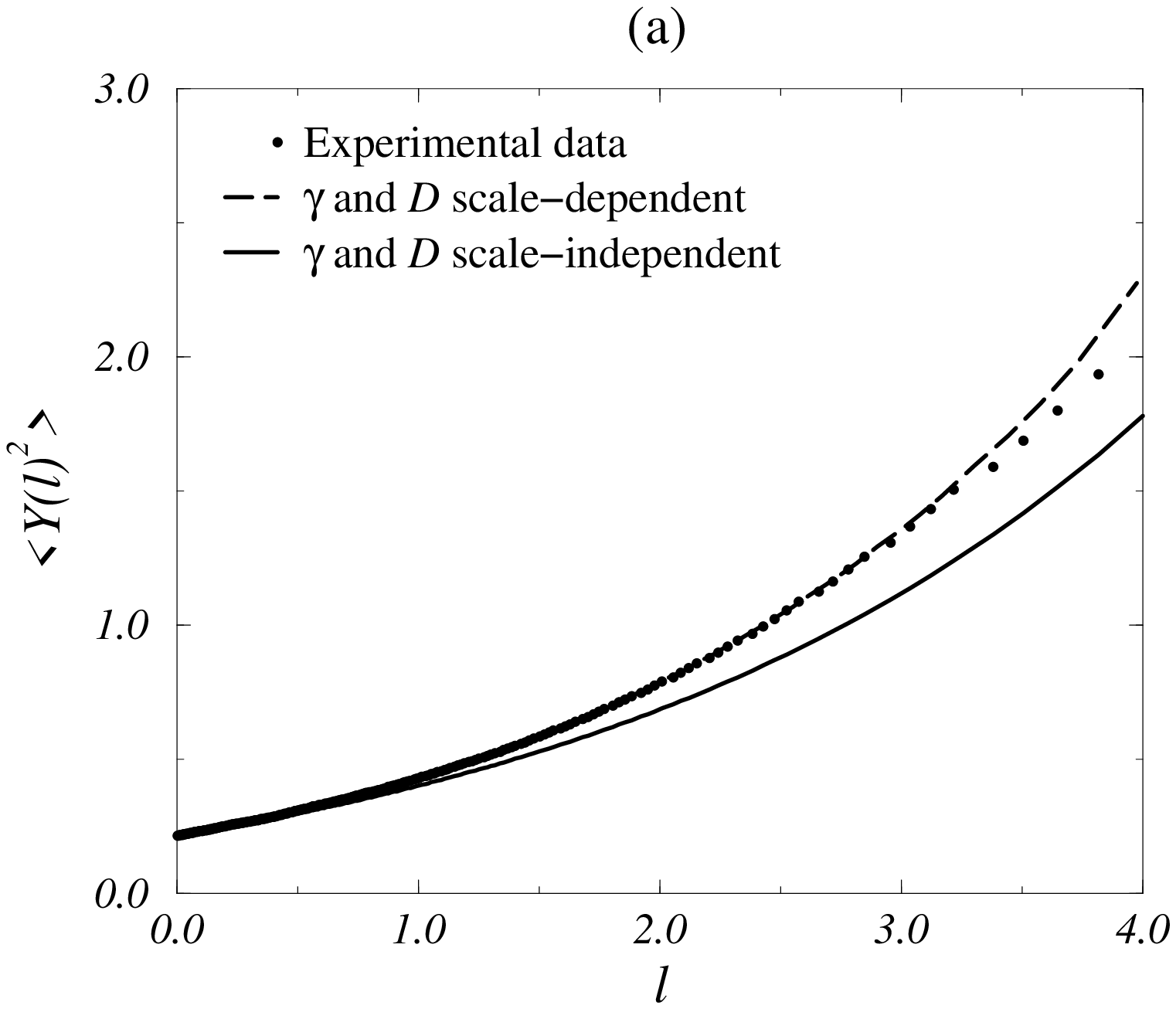}
\hspace{-1.5cm}
\epsfxsize 10cm
\epsffile{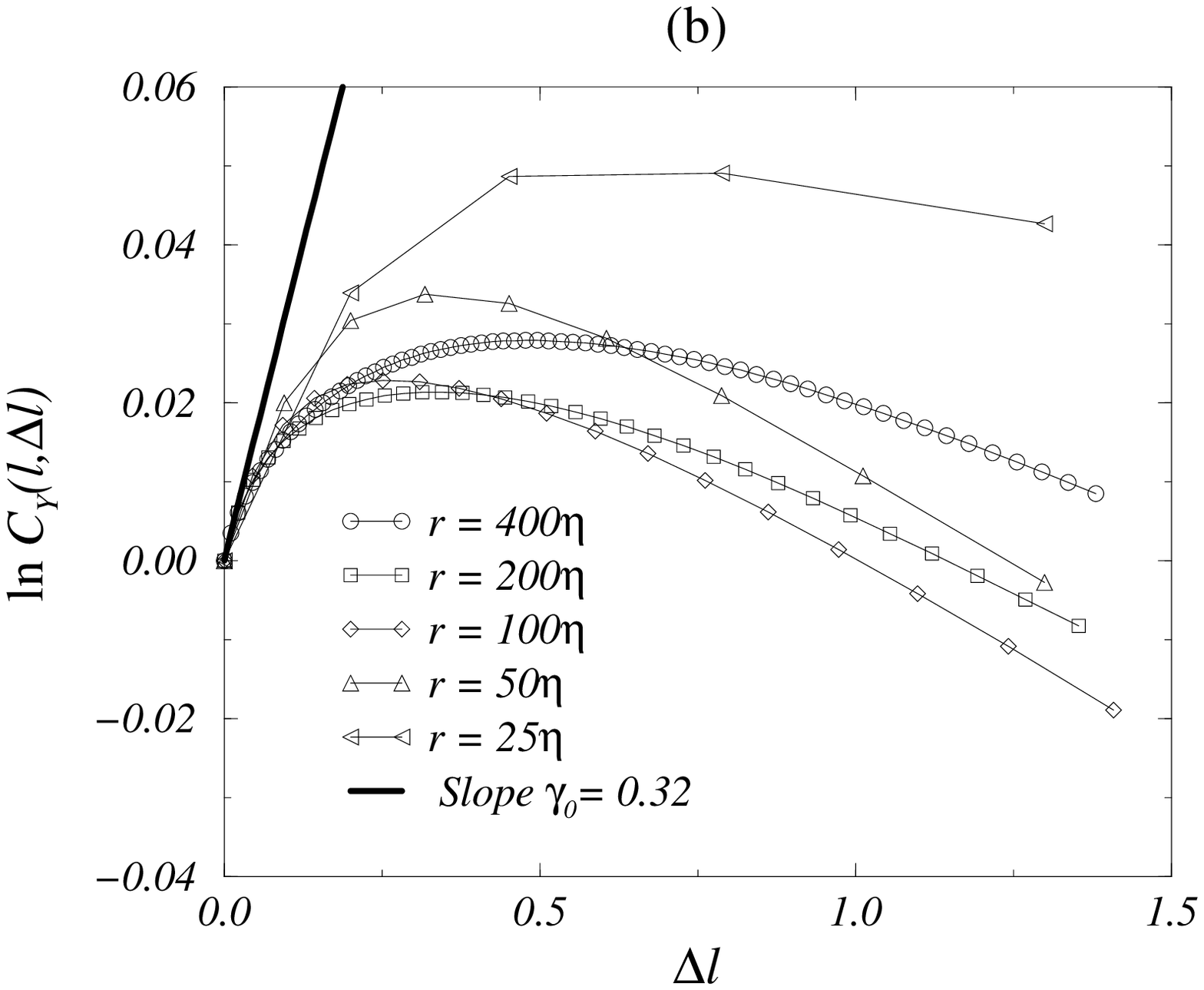}
}
\vspace{-0.5cm}
\caption{Graph (a): variance of $Y(l)$. The statistical error on
experimental data is smaller than the width of symbols. The prediction
obtained in the case of scale-dependent coefficients (Eq.~(\ref{varY2}),
dashed line) is indistinguishable from experimental data in the interval
$0 \le l \le 3$. See text for the numerical values used in the 
calculation of $\langle Y(l)^2 \rangle$. Graph (b): lin-log plot of 
the scale autocorrelation functions $C_Y(l,\Delta l)$ vs. the
scale difference $\Delta l$, for scales $l$ ranging from
$l = 25 \eta$ to $l = 400 \eta$.
A straight line with slope $\gamma_0$ is also shown for comparison
with the behaviour expected in the vicinity of $\Delta l = 0$ 
in the case of scale-dependent drift and diffusion coefficients
(Eq.~(\ref{corY2})).}
\label{fig-pred}
\end{figure}

In Fig.~\ref{fig-pred}.a, we compare the scale dependence 
of $\langle Y(l)^2 \rangle$, as obtained from experimental
data, with predictions (\ref{varY}) and (\ref{varY2}).
The initial value is set to $\langle Y(0)^2 \rangle = 0.23$.
We use the following numerical values: 
$\gamma = 0.21$ (Eq.~(\ref{varY})),
$\gamma_0 = 0.32$, $\gamma_1 = 0.025$, $D_0 = 0.01$, $\delta = 0.40$
(Eq.~(\ref{varY2})).
We find that Eq.~(\ref{varY2}), {\it not} Eq.~(\ref{varY}), 
fits experimental data extremely well.
A small deviation is observed only for scales smaller than
$r = 25 \eta$ ($l \ge 3.0$), where the diffusion coefficient
$D(r)$ deviates from the scaling law (\ref{valD}) 
(see Fig.~\ref{fig-D}). This first result suggests that
approximating the drift and diffusion coefficients to
a constant value is inappropriate.

Fig.~\ref{fig-pred}.b shows the measured scale autocorrelation functions 
$C_Y(l, \Delta l)$. Again, the linear growth predicted by
Eq.~(\ref{corY}) is not observed. However, the following features 
of the scale-dependence of $C_Y(l, \Delta l)$ are reproduced 
by Eq.~(\ref{corY2}) at a qualitative level. The autocorrelation
function 
is a function of both $l$ and $\Delta l$. For all (fixed) values of $l$,
its logarithm first grows with $\Delta l$ before reaching a
maximum and eventually decaying. The tangent at $\Delta l = 0$ is
smaller than $\gamma_0 = 0.32$. However, quantitative agreement is still 
lacking, for instance concerning the location of the maximum 
of $C_Y(l, \Delta l)$, as well as the exact dependence
in $\Delta l$. 
We will show in Sec.~\ref{sec-Markov} that,
as one would now expect, a $\delta$-function is indeed a good
approximation 
of the autocorrelation function of the noise $C_{\xi}(l, \Delta l)$, 
as shown by comparing the correlation scale of $C_{\xi}$ 
to the evolution scale of $C_Y$. 
We therefore believe that the quantitative discrepancy 
observed between the prediction
(\ref{corY2}) and experimental data should be attributed to the 
approximation made when fitting $\gamma(r)$ by a logarithmic
function of scale. In particular, the oscillations apparent
in the inset of Fig.~\ref{fig-gamma} for large $l$ do not
seem to be statistical fluctuations, and should therefore be taken
into account when computing $C_Y(l, \Delta l)$.

In conclusion, $\gamma(l)$ and $D(l)$ should not be approximated
to a constant value. Further, the functional dependence given
by Eqs.~(\ref{valgam}) and (\ref{valD}) is consistent with
experimentally obtained two-point correlators of the process
$Y(l)$, qualitatively only for the autocorrelation function
$C_Y(l, \Delta l)$, and quantitatively for the variance 
$\langle Y(l)^2 \rangle$.

\subsection{Discretisation scheme}
\label{sec-Langevin-disc}

In the following, we will consider the Langevin equation
(\ref{Langevin2}),
where $\gamma(l)$ and $D(l)$ are defined by Eqs.~(\ref{mesgam2}) and
(\ref{mesD}) respectively. Provided that the noise term $\xi(t)$ is Gaussian, 
stationary, and delta-correlated, this equation is formally 
equivalent to a Fokker-Planck equation such as Eq.~(\ref{FP}),
with scale-dependent drift and diffusion coefficients.
As before, this Fokker-Planck equation admits a Gaussian
solution, the variance of which is given by Eq.~(\ref{varY2}.

The stochastic differential equation (\ref{Langevin2}) may be
discretised
according to the following first-order scheme \cite{Risken}:
\begin{equation}
\label{Ldiscrete}
Y\left(l+\delta l \right) - Y(l) = \gamma(l) \; Y(l) \; \delta l + 
\sqrt{2 \; D(l) \; \delta l} \; \xi(l).
\end{equation}
In the following, we will use Eq.~(\ref{Ldiscrete}) as an operational
definition of the driving random force:
\begin{equation}
\label{bruit}
\xi(l) \equiv {1 \over \sqrt{2 \; D(l) \; \delta l}}  \left(
Y(l+\delta l) - Y(l) - \gamma(l) \; Y(l) \; \delta l \right).
\end{equation}
By construction, $\xi(l)$ is a centered stochastic variable
at all scales $l$: $\langle \xi(l) \rangle = 0$,
with unit variance $\langle \xi(l)^2 \rangle = 1$.

In Secs.~\ref{sec-Markov} and \ref{sec-stat}, we will 
respectively test the Markov and Gaussian nature of the 
stochastic process defined by  Eq.~(\ref{Langevin2}).

\section{A Markov process}
\label{sec-Markov}

In this section, we further investigate the validity of
Eq.~(\ref{Langevin2}) by turning to the dynamics of the process,
and in particular to two-point scale autocorrelation functions.

Due to the viscosity of the fluid, the velocity field 
of a turbulent flow remains differentiable. 
The energy cascade cannot be perfectly 
represented by a Markov process: the process $\xi(l)$ must also be 
differentiable, and can therefore not be $\delta$-correlated.
Its (normalised) two-point scale autocorrelation function 
is defined as:
\begin{equation}
C_{\xi}(l,l') = \left\langle \xi(l) \; \xi(l') \right\rangle.
\end{equation}

\begin{figure}[thb]
\vspace{-0.5cm}
\centerline{
\epsfxsize 10cm
\epsffile{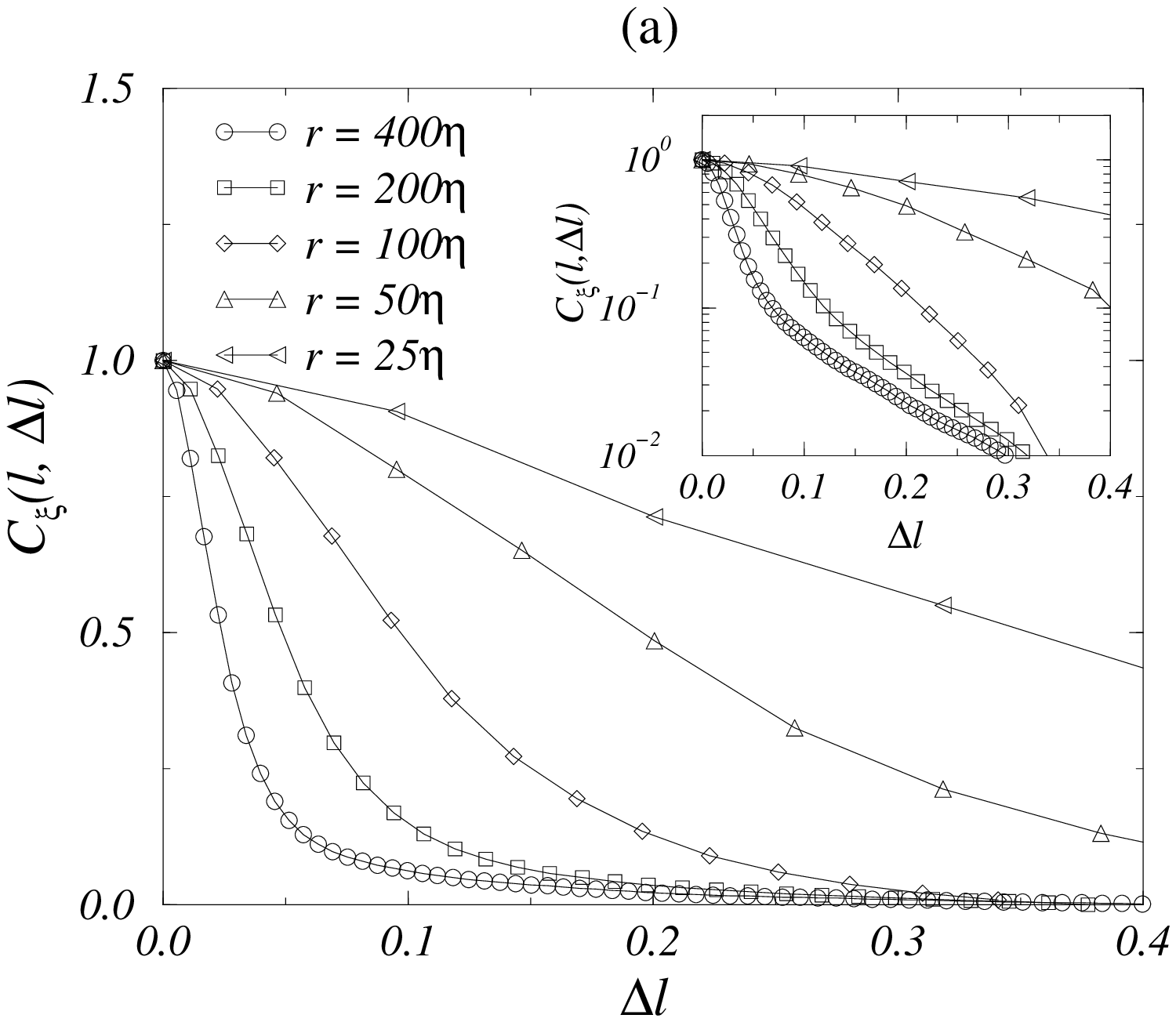}
\hspace{-1.5cm}
\epsfxsize 10cm
\epsffile{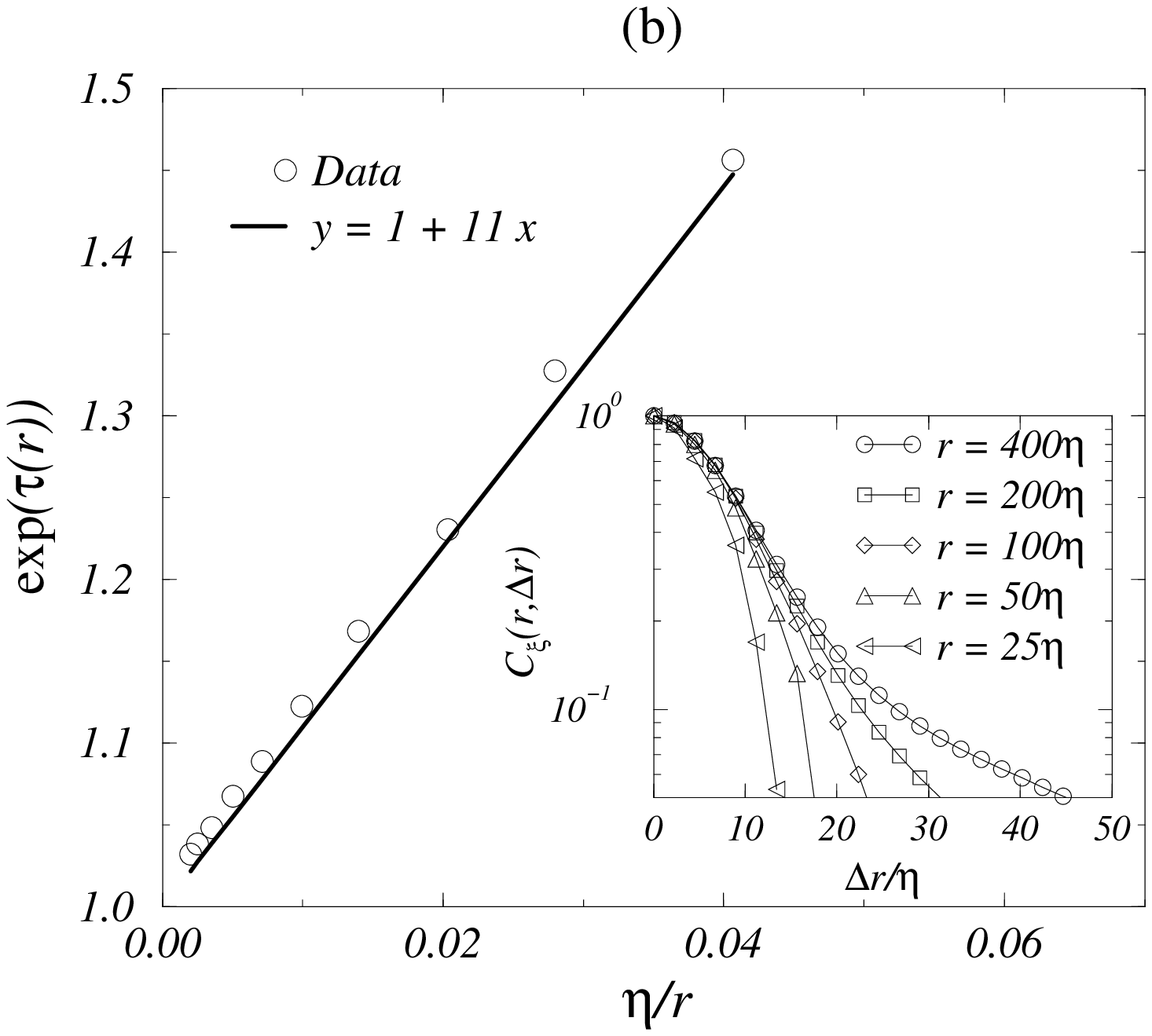}
}
\vspace{-0.5cm}
\caption{Graph (a): scale autocorrelation function $C_{\xi}(l,\Delta l)$ 
of the random force $\xi(l)$ for scales $r = L \exp(-l)$ ranging
between $r = 25 \eta$ and $r = 400 \eta$.
The statistical error is smaller than the width of symbols.
Inset: lin-log plot of the same curves.
Graph(b): scale-dependence of the autocorrelation scale $\tau(r)$.
Inset: same graph as (a), represented as a function of the physical 
scale $r$.}
\label{fig-cor-xi}
\end{figure}

Fig.~\ref{fig-cor-xi}.a shows that the scale dependence of
$C_{\xi}(l,l')$ 
cannot be reduced to the form $C_{\xi}(l'-l)$. The process
is thus instationary in terms of the scale variable $l$.
One may however write without loss of generality:
\begin{equation}
\label{inst}
C_{\xi}(l,l') = C_{\xi}(l,\Delta l),
\end{equation}
where $\Delta l = l'-l$ is set positive by convention.
Further, correlations of the random force differ from the ideal
case (Eq.~(\ref{noise2})): the autocorrelation scale
$\tau(l)$ is non-zero, its numerical value depends on scale $l$.
Even though the scale autocorrelation function 
decays approximately exponentially over roughly one decade,
we choose to evaluate the characteristic 
correlation scale $\tau(l)$ from the expression :
\begin{equation}
\label{tauint}
\tau(l) = \int_{0}^{\infty} C_{\xi}(l, \Delta l) \, {\rm d}\Delta l.
\end{equation}
We checked that this particular choice does not affect our conclusions.

The autocorrelation scale $\tau(l)$ 
is a quantitative measure of the departure from the Markov
approximation.
At a given scale $l$, the scale $\tau(l)$ marks the limit
below which the stochastic cascade process becomes smooth.
In this sense, the (measurable) scale $\tau(l)$ is the elementary 
step of the energy cascade. 

Eq.~(\ref{tauint}) yields numerical values ranging between 
$\tau = 0.37$ when $r = 25 \eta$ and $\tau = 0.03$ when 
$r = L = 500 \eta$.
Fig.~\ref{fig-cor-xi}.b shows that the scale-dependence 
of $\tau(l)$ may be fitted by the expression: 
\begin{equation}
\label{tau}
\tau(r) = \ln \left({r+r_0 \over r}\right),
\end{equation}
for a numerical value of $r_0$ close to $11 \; \eta$.
In other words, the instationarity observed at the level of
autocorrelation
functions (Fig.~\ref{fig-cor-xi}.a) is mostly due to the 
change of variable $l = \ln(L/r)$.
This is confirmed by the insert of Fig.~\ref{fig-cor-xi}.b,
where the autocorrelation functions $C_{\xi}$ are
plotted with respect to the physical scale difference $\Delta r$:
the first decade of decay is characterised at all scales by 
a slope roughly equal to $1/r_0$.
In terms of the scale variable $r$, the elementary
cascade step may simply be defined as $r_0$,
of the order of Kolmogorov's scale $\eta$. 

The characteristic evolution scale of the random process $Y(l)$ 
at scale $l$ is given by the inverse drift coefficient $1/\gamma(l)$ 
(cf. Eq.~(\ref{Langevin2})). The product $\gamma(r) \tau(r)$ is a 
monotonically decreasing function of scale $r$, with 
$\gamma(25 \eta) \tau(25 \eta) = 0.06$ and
$\gamma(L) \tau(L) = 0.01$. 
In the range of scales $25 \eta \le r \le 500 \eta$ at least, one finds
that:
\begin{equation}
\label{appmarkov}
\tau(l) \ll {1 \over \gamma(l)},
\end{equation}
by more than one order of magnitude. Although the correlation scale
$\tau(l)$ is non-zero, the random process $Y(l)$ is therefore
effectively Markov.
Moreover, the product $\gamma(r) \tau(r)$, as defined
by Eqs.~(\ref{valgam}) and (\ref{tau}), admits an absolute
maximum close to $0.1$ in the vicinity of $r \sim 5 \eta$.
Assuming that Eqs.~(\ref{valgam}) and (\ref{tau}) faithfully 
describe the scale dependence of $\gamma(l)$ and $\tau(l)$ down to 
the smallest scales, this implies that the energy cascade is always
a Markov stochastic process for scales larger than $\eta$.

\section{A non-gaussian random force}
\label{sec-stat}

In this section, we investigate the validity of
Eq.~(\ref{noise3}) by evaluating the probability distribution function
$P(\xi,l)$ of the random force $\xi$ at scale $l$.

\subsection{A typical realisation of the process $\xi(l)$}
\label{sec-stat-traj}

\begin{figure}[thb]
\vspace{-0.5cm}
\centerline{
\epsfxsize 10cm
\epsffile{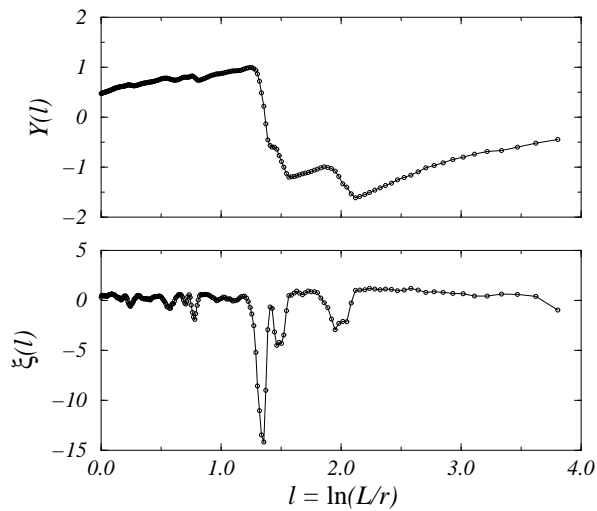}
}
\vspace{-0.5cm}
\caption{A typical random trajectory of the stochastic 
processes $Y(l)$ and $\xi(l)$. Note that $Y(0)$
is not equal to zero.}
\label{fig-traj}
\end{figure}

A typical realisation of the random processes $Y(l)$ 
and $\xi(l)$ is presented in Fig.~\ref{fig-traj}.
The dynamics of 
$Y(l)$ is generally dominated 
by the deterministic part of Eq.~(\ref{Langevin2}).
Long periods of quasi-exponential growth 
controlled by the drift coefficient $\gamma > 0$ are perturbed by small 
deviations due to the random force $\xi(l)$. This regime is 
interrupted infrequently by sharp drops, 
corresponding to large negative excursions of the random force.
This suggests that the probability distribution function 
of $\xi$ is strongly skewed, in contrast with the Gaussian prediction.
Accessible values of $\xi(l)$ also seem to be bounded
from above.

\subsection{Probability distribution function of the random force}
\label{sec-stat-pdf}

\begin{figure}[thb]
\vspace{-0.5cm}
\centerline{
\epsfxsize 10cm
\epsffile{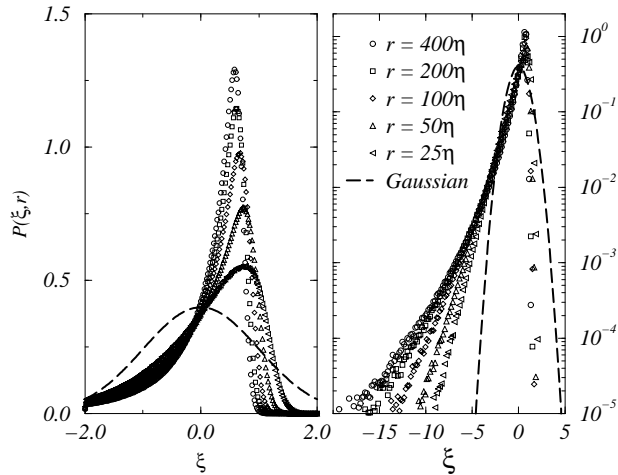}
}
\vspace{-0.5cm}
\caption{Probability distribution functions $P(\xi,r)$
of the random force $\xi$ for scales $r$
ranging from $r = 25 \eta$ to $r = 400 \eta$ in lin-lin
(left-hand side) and lin-log (right-hand side) plots. The dashed line
is drawn for comparison with the Gaussian prediction derived from
the Fokker-Planck model.
}
\label{fig-pdfxi}
\end{figure}

As shown in Fig.~\ref{fig-pdfxi}, the probability distribution function 
$P(\xi,r)$ of the random force $\xi$ at scale $r$ is markedly 
different from the Gaussian prediction (Eq.~(\ref{noise3})):
the skewness factor of $P(\xi,r)$ is indeed negative.
The shape of $P(\xi,r)$ depends on scale $r$: this is consistent 
with the instationary behaviour of autocorrelation 
functions (cf. Sec.~\ref{sec-Markov}).
Even though Gaussian-like, rapid decay of $P(\xi,r)$
is observed for $\xi > 0$, the probability of large negative deviations
is much larger than that predicted for a Gaussian process. 
This long tail can be fitted with reasonable accuracy either by a
log-normal or by a stretched-exponential functional form.

The Gaussian-like decay of $P(\xi)$ for positive values of
the random force may be understood as follows. 
The definition 
of $\epsilon_r$ as a sum of non-negative quantities
(Eq.~(\ref{epsilon}))
implies that:
\begin{equation}
\label{const1}
r' \; \epsilon_{r'} < r \; \epsilon_{r}, \; \forall \; r' < r.
\end{equation}
The process $X(l)$ can only adopt values such that:
\begin{equation}
\label{const2}
X(l') - X(l)  < l' - l, \; \forall \; l' > l.
\end{equation}
Taking the appropriate limit, one obtains:
\begin{equation}
\label{const3}
\dot{Y}(l) \le 1 - {{\rm d} \over {\rm d}l} 
\langle X(l) \rangle.
\end{equation}
The random force $\xi$ is defined as the difference of the
two random variables
$\dot{Y}$ and $Y$, where $\dot{Y}$ admits an upper bound and the
fluctuations of $Y$ are nearly Gaussian: the tail of $P(\xi,l)$
is thus expected to be Gaussian-like for positive 
values of the random force $\xi$. 

At a qualitative level, the asymmetry of the probability 
distribution function of
$\xi$ is readily understood if one remembers that
$P(Y,r)$ is also asymmetrical, as seen in Fig.~\ref{fig-pdfy}.
The solution of a linear stochastic evolution equation
with Gaussian random driving is symmetrical.
Conversely, the skewness of the variable $Y(l)$ points at the
necessity of corrections to the Fokker-Planck model.
The dominant one is expected to stem from the third-order 
coefficient $D_3$. Non-zero $n$-th order coefficients 
$M_n$ translate into deviations of the $n$-point correlators 
of the random force $\xi(l)$ from their Gaussian form. 
In particular, a non-zero
value of $D_3$ translates into a non-zero skewness, as observed here. 
The asymmetry of $P(Y,r)$ is weak, as quantified by the
small value of the parameter $|D_3(l)|/D(l)^{3/2}$,
which we found smaller than $10^{-1}$ at all scales.

Intermittency of fully-developed turbulent flows is
often characterised by the presence of large excursions of the 
(positive) local dissipation rate $\epsilon(x) = 15 \nu (dv/dx)^2$ 
far above its mean value. We believe that these rare events 
also correspond to large negative deviations of the process $\xi(l)$.
The dissipation rate $\epsilon_r(x)$ at location $x$ is defined as the
spatial average of $\epsilon(x)$ over an interval of length $r$ 
(Eq.~(\ref{epsilon})). The location $x$ being fixed, $\epsilon_r(x)$
is a monotonously decreasing function of
scale $r$ when the integral of $\epsilon(x)$ over an interval
of length $r$ grows more slowly than $r$, i.e. when fluctuations
of $\epsilon(x)$ are weak enough. This case corresponds to
the periods of quasi-exponential growth of $Y(l)$ in Fig.\ref{fig-traj}.
Assume now that a sharp increase of $\epsilon(x)$ takes place
close to $x^* = x + r^*/2$. Close to $r^*$, the value of 
$\epsilon_r(x)$ will increase sharply: 
$\epsilon_{r^* + \delta r}(x) \gg \epsilon_{r^* - \delta r}(x)$,
corresponding to a large positive value of the scale derivative 
$\partial \epsilon_r(x) / \partial r_{|r = r^*}$. 
Since the logarithmic scale 
$l$ is a monotonously decreasing function of the physical scale $r$,
this shows that intermittent bursts of $\epsilon(x)$ do indeed
correspond to 
sharp and localised drops of $Y(l)$. In other words, the asymmetry of 
$P(\xi, r)$ expresses the intermittent character of the local 
dissipation field.

However, the definition of $\epsilon_r$ as the averaged dissipation rate 
over a segment of length $r$ is to some extent arbitrary. 
The amplitude of intermittent events, as quantified by 
the value of $\xi(l^*)$ close to a large deviation of the 
local dissipation rate $\epsilon(x^*)$, may depend sharply
on the precise definition of the averaging procedure which 
leads to $\epsilon_r(x)$. In this sense, we conjecture that
the precise numerical value of, e.g., the skewness coefficient of 
$P(\xi,r)$ is in fact not relevant to the description of the physical 
cascade process, since it may depend on arbitrary features 
of the analysis method, such as the choice of an averaging window 
in Eq.~(\ref{epsilon}).

\section{Discussion}
\label{sec-conc}

Following \cite{Naert97}, we have found additional evidence 
that the energy cascade of a turbulent flow 
at high Reynolds number is well described by 
a continuous stochastic process of Ornstein-Uhlenbeck type.
We also clarified the validity of a number of approximations 
made in \cite{Naert97}.

Defining the stochastic process $Y(l)$ as the centered variable
$\ln \epsilon_r - \langle \ln \epsilon_r \rangle$
at scale $l = \ln(L/r)$, the relevant Langevin equation 
is linear, and reads:
\begin{equation}
\label{Langevin3}
{dY \over dl}(l) = \gamma(l) Y(l) + \sqrt{2D(l)} \; \xi(l).
\end{equation}
We found that the scale-dependence
of the drift and diffusion coefficients
$\gamma(l)$ and $D(l)$, as evaluated directly from 
experimental data by methods independent from those
used in \cite{Naert97}, is well described by the
following functional forms:
\begin{equation}
\label{rescoefs}
\begin{array}{lcl}
\gamma(l) &=& \gamma_0 - \gamma_1 \; l,\\
D(l) &=& D_0 \; e^{2 \; \delta \; l},
\end{array}
\end{equation}
where $\gamma_0 = 0.32 \pm 0.03$, 
$\gamma_1 = 0.025 \pm 0.003$,
$D_0 = 0.01 \pm 0.005$,
and $\delta = 0.40 \pm 0.01$ are positive constants.
The (exact) solution of Eqs.~(\ref{Langevin3})-(\ref{rescoefs})
is in good agreement with experimental data at the level
of two-point correlators of the stochastic variable $Y(l)$.
Further, we have shown that the drift and diffusion 
coefficients cannot be approximated by constant values
if one wants to preserve this agreement.

The main characteristics of the random force $\xi(l)$
have been determined directly from experimental data.
We have checked that the process is Markov at all scales $l$, since
scale autocorrelation functions of the random force 
decay rapidly on a characteristic autocorrelation 
scale $\tau(l)$ much smaller than the typical evolution scale
$1/\gamma(l)$ of the process. This is perhaps our most important 
result: the Langevin equation (\ref{Langevin3}) defines a driving 
random force $\xi(l)$ which respects the Markov hypothesis.
Previous work \cite{Naert97} only provided evidence for 
the validity of a necessary condition for
the additive process (\ref{Langevin2}) to be Markov. 
The autocorrelation scale $\tau(l)$ is the elementary step of the
cascade
process. It depends on the physical scale $r$ as
$\tau(l) = \ln \left({(r+r_0) / r}\right)$, where $r_0$
is of the order of Kolmogorov's scale $\eta$.

Finally, our analysis of experimental data
shows that the probability distribution function of 
the random force $\xi(l)$ is strongly non-Gaussian,
and exhibits a long tail for negative noise
(see Fig.~\ref{fig-pdfy}). 
These rare, but intense deviations are correlated with 
large positive values of the scale derivative 
$\partial \epsilon_r/ \partial r$.
For a linear equation such as (\ref{Langevin2}),
the non-zero skewness of $P(\xi,l)$ translates
into an asymmetrical solution $P(Y,l)$, in qualitative agreement 
with observation.
This deviation from the simple Ornstein-Uhlenbeck picture is
equivalent to a correction to log-normal statistics for the energy 
dissipation rate $\epsilon_r$. Within the Fokker-Planck 
description, the corresponding
leading-order correction is the third-order term
of the Kramers-Moyal expansion (Eq.~(\ref{KM})), 
with a non-zero, scale-dependent coefficient $D_3(l)$.
This correction is weak, as quantified by the
value of the ratio $|D_3(l)| / D^{3/2}$, smaller than $10^{-1}$ 
at all scales. This correction is made at the expense of the solvability
of our model. 

Two comments are in order. First, this stochastic description 
of the energy cascade (including the Markov property)
applies to all scales, from the integral scale
$L$ down to the dissipation scale $\eta$.
The distinction commonly observed between 
dissipative and inertial sub-range statistics of, e.g.,
turbulent velocity and passive scalar density fields \cite{book},
appears to be irrelevant in the case of the energy dissipation rate.
Second, all known multiplicative models of
the cascade use, at least to our knowledge, 
a scale-independent discretisation step such as $\ln 2$.
Although natural from a mathematical viewpoint, this
choice overlooks the scale-dependence of $\tau(l)$
observed here, and may thus not be appropriate on physical grounds.
The choice of the discretization step of a discrete stochastic
model of the energy cascade is not meaningless.

We would now like to emphasise that
observations reported in this work stem from the analysis of one 
turbulent velocity signal, recorded
in a particular realisation of a jet flow at a given value of 
the Reynolds number. One would of course like to know
to what extent the Langevin equation (\ref{Langevin3}) 
is a universal description of the energy cascade.
Preliminary studies have already shown that
the statistical properties of the processes $\xi(l)$ and $Y(l)$ 
are qualitatively similar to those discussed here
for other values of the Reynolds number \cite{encours}.
Further, a number of adimensional quantifiers of the energy 
cascade have been introduced at a rather formal level ($\gamma(l)$,
$D(l)$, etc.). One would like to understand their physical meaning,
as indicated by, e.g., a possible dependence on the physical properties 
of the fluid and the Reynolds number, and in particular to 
elucidate their behaviour in the limit of large $Re$. 
Conversely, it is also essential to check whether some of
our results, such as the numerical value of, say, the skewness
of $P(\xi,r)$, may not be artefacts of the data processing method. 
These important points are currently being considered,
thanks to the analysis of the velocity fields of other turbulent flows.

Interestingly, the existence of a stochastic evolution equation 
in scale allows, at least in principle, to reconstruct 
by simple integration the small scale statistics of the
dissipation field from spatial fluctuations observed at
larger scale. We therefore believe that this description 
of the energy cascade process may prove useful, as a model of 
the small scale fluctuations, in the context of 
large eddy simulations where only the larger
scales of a flow are usually resolved.

\ack
The authors would like to thank Yoshiki Kuramoto for his support
as well as for many insightful comments, Jean-Fran\c{c}ois
Pinton for a critical reading of the manuscript, 
and Bernard Castaing for a useful discussion.
The warm hospitality of Kyoto university, Tohoku university
and Ecole Normale Sup\'erieure de Lyon is gratefully acknowledged.

\end{document}